\documentclass[a4paper,12pt]{article}
\usepackage{amssymb,amsfonts,enumerate}
\usepackage{graphicx,array}

\newlength{\defbaselineskip}
\begin{document}

\title{ Measurement and Macroscopicity: Overcoming Conceptual Imprecision in Quantum Measurement Theory}

\author{Gregg Jaeger\\
             	Quantum Communication and Measurement Laboratory,\\ 
	         Department of Electrical and Computer Engineering\\
             	and Division of Natural Science and Mathematics,\\ Boston University, Boston, MA \\
}\maketitle
  
\begin{abstract}
John S. Bell is well known for the result now referred to simply as ``Bell's theorem,'' which removed from serious consideration by physics of local hidden-variable theories. Under these circumstances, if quantum theory is to serve as a truly {\em fundamental} theory, conceptual precision in its interpretation is not only even more desirable but paramount. John Bell was accordingly concerned about what he viewed as conceptual imprecision, from the physical point of view, in the standard approaches to the theory. He saw this as most acute in the case of their treatment of {\em measurement at the level of principle}. Bell pointed out that conceptual imprecision is reflected in the terminology of the theory, a great deal of which he deemed worthy of banishment from discussions of principle. For him, it corresponded to a set of what he saw as vague and, in some instances, outright destructive concepts. Here, I consider this critique of standard quantum measurement theory and some alternative treatments wherein he saw greater conceptual precision, and make further suggestions as to how to proceed along the lines he advocated.
\end{abstract}

\vfill\eject

\section{Introduction}

John S. Bell is well known for the result now referred to simply as ``Bell's theorem,'' which removed from consideration the class of so-called local hidden-variable theories which at the time of its publishing appeared to be the most natural class of theories among those that would render quantum mechanics a form of statistical mechanics. If, as this and other results suggest, quantum theory is to serve as a truly {\em fundamental} theory, conceptual precision in its interpretation is not only desirable but paramount. John Bell was accordingly concerned about what he viewed as conceptual imprecision, from the physical point of view, in the standard approaches to the theory. He saw this as most acute in the case of their treatment of {\em measurement at the level of principle}. His concerns were strongly expressed in one of his last articles,  ``Against Measurement.'' This item was published in 1990 in a volume of proceedings of the 1989 Erice meeting ``Sixty-Two Years of Uncertainty,'' during which it was my pleasure to meet and eat with Bell, and to listen to him present this paper. He pointed out that this conceptual imprecision is reflected in the {\em terminology} of the foundations of quantum theory, a great deal of which he  deemed worthy of banishment from discussions of principle, because it corresponds to a set of what he saw as vague and, in some instances, outright destructive concepts. His concern was not one regarding the mathematics so much as regarding basic concepts used in contemporary quantum physics, which he viewed as failing to satisfy the needs of natural philosophy and so of physics, despite their apparent practical adequacy. Here, I consider Bell's critique of standard quantum measurement theory and some alternative treatments wherein he saw greater conceptual precision, and make further suggestions as to how to improve conceptual precision, as he advocated.

That the source of difficulties is to be understood specifically as a problem of {\em imprecision of physical concepts} which stands in the way of the achievement of an exact fundamental mechanical theory is pointed out at the outset of ``Against measurement'' (AM): Bell wished to make it clear ``at once that it is not mathematical precision, but physical'' that caused him such great concern. As he saw it, physics should have had by the time of its writing  
``an exact formulation of a serious part of  mechanics,'' where by  exact he meant ``only that the theory should be fully formulated in mathematical terms, with nothing left to the discretion of the theoretical physicist,'' with nonrelativistic `particle' quantum mechanics and that of the electromagnetic field constituting a sufficiently ``serious part'' (Bell, 1991a).
Bell also made it immediately clear that he saw physics as part of the long tradition of  {\em natural philosophy}, and that his concerns about physical precision are, in effect, concerns regarding the {\em precision of concepts} of natural philosophy. 

In the analysis offered in AM,  a key distinction, made with regard to theoretical treatments compatible with experimental data obtained, is that between those sufficiently precise to be accepted as fundamental physics and those good enough ``for all practical purposes,'' for which he supplied the memorable acronym `FAPP' (which, following his usage, continues to be employed in this sense now, two and one half decades later). 
Bell's exploration begins by pointing out that there is a lack of precision in the traditional, ``proper treatments'' one finds published in respectable
and frequently consulted sources. 
He recalls and answers the often-asked rhetorical question of why one should bother making quantum mechanics more precise
than it already is: ``Why not look it up in a good book? But {\em which} good book? Usually the good 
unproblematic formulation is still in the head of the person in question\ldots For the good books known to me are not much concerned 
with physical precision.'' His verdict on those various available treatments that are set firmly within quantum theory involving the standard, unmodified 
dynamical laws is that they are useful for the practical prediction of the statistics to be found in experiments but fall far short of what physics ought 
to be  {\em at the level of principle}: ``The orthodox approaches, whether the authors think they have made derivations or assumptions, are just fine FAPP'' but ultimately fail to fully describe the physical world  (Bell, 1991).

Bell provides a lengthy laundry list of standard quantum physical terms, reflection upon which shows that the lack of physical precision in the then current thinking---from which, it should be noted out, we have yet to significantly advance---is due to {\em conceptual} imprecision, and
suggests that physics reject a considerable amount of this standard terminology.
\begin{quote}
``Here are some words which, however legitimate and necessary in application, have no place in a {\em formulation} with any pretension
of physical precision: {\em system, apparatus, environment, microscopic, macroscopic, reversible, irreversible, observable, information,
measurement.} The concepts of `system', `apparatus', `environment', immediately imply an artificial division of the world, and an intention 
to neglect, or take only schematic account of, the interaction across the split. The notions of `microscopic' and `macroscopic' defy precise
definition. So also do the notions of `reversible' and `irreversible'. Einstein said that it is theory which decides what is `observable'. I think 
he was right --- `observation' is a complicated and theory-laden business. Then that notion should not appear in the {\em formulation} of 
fundamental theory.'' (Bell, 1991)
\end{quote}

The ages old philosophical question of the relationship of observation to reality is relevant to the issues discussed in the article, but is itself not engaged in the text in any detail, beyond a general rejection of subjectivism. It suffices here to note that in his writings generally, Bell sides with realism and in AM refers positively to the views of one of its great physicist-champions on this issue.
His primary concern instead is more specifically the relationship of {\em physical theory} to reality, and his position is that, at a minimum, physical theory should should explain
to the physicist what can and cannot be measured, something given by the answer to the question of how measurements are and can be made without themselves being considered {\em fundamental} to physical theory.

\section{The Negative Influence of Inappropriate Terminology}

As Bell saw it in AM, ``On this list of bad words from good books, the worst of all is `measurement'.'' Again, he does not reject the term
in general, particularly not its use in practice and mentioned that in the command ``measure the mass and width of the Z 
boson" as an example of acceptable use of it. He objected, rather, and most specifically to ``its use in the fundamental interpretive rules of quantum mechanics.'' 
Bell considered the problems as arising through the use of `measurement' in the {\em foundations} of quantum theory,  commenting that, when reading of Dirac's ``good book'' {\em Quantum mechanics}, one gets the sense that 
``the theory is exclusively concerned about `the results of measurement', and has nothing to say about anything else'' but, he asks rhetorically, ``What qualifies some 
physical systems to play the role of `measurer'?'' (Bell, 1991). 

Bell suggests that the role of the notion of measurement be taken over 
by the more neutral term `experiment': ``Even in a low-brow practical account, I think it would be good to replace the word `measurement', in the formulation, by the word `experiment'. For the latter word is altogether less misleading.'' But, this move also has limitations.
\begin{quote}
``However, the idea that quantum mechanics, our most fundamental
physical theory, is exclusively about the results of experiments would remain disappointing. \ldots To restrict quantum mechanics to be 
exclusively about piddling laboratory operations is to betray the great enterprise [of natural philosophy]. A serious formulation will
not exclude the big world outside the laboratory.'' (Bell, 1991)
\end{quote}
He finds use of the term `experiment' tolerable in the formulation of quantum mechanics but best avoided if possible.
The term `measurement,' however, is ``entirely inappropriate.'' 
He makes two specific charges against the term:
\begin{quote}
``The first charge against `measurement', in the fundamental axioms of quantum mechanics, is that it anchors there the shifty split of the
world into `system' and `apparatus'. A second charge is that the word comes loaded with meaning from everyday life, meaning which is
entirely inappropriate in the quantum context \ldots In other contexts, physicists have been able to take words from everyday language and
use them as technical terms with no great harm done \ldots Would that it were so with `measurement'. But in fact the word has had such
a damaging effect on the discussion, that I think it should now be banned altogether in quantum mechanics.'' (Bell, 1991)
\end{quote}

Beyond the general difficulties of its use of this most problematic term, Bell sees the traditional, ``orthodox'' treatment of  measurement-like processes as reinforcing the imprecision of the enterprise of quantum natural philosophy in several ways. He also objects to any distinction between systems based on imprecise reference to {\em physical scale}, where the term `macroscopic' is brought into play.
\begin{quote}
``The kinematics of the world, in this orthodox picture [with probabilities of obtaining outcomes], is given by a wavefunction (maybe more
than one?) for the quantum part, and classical variables --- variables that {\em have} values --- for the classical part: ($\Psi(t,q,\ldots), X(t\ldots\ldots))$. The $X$s are somehow macroscopic. This is not spelled out very explicitly. The dynamics is not very precisely formulated either.'' (Bell, 1991)
\end{quote}
`Macroscopic' is another term set for banning. Bell had expressed concern with that term in previous years as well, but he had previously seen a sharpening of the concept as still viable; he had commented, for example, that in regard to the 
`EPR correlations' which
violated his inequality that he had ``very little understanding of the position of\ldots Bohr,'' which depended on restrictions on what is to be considered possible in measurements and made use of the term (Bell, 1981a). 
In his 1981 article entitled ``Bertlmann's socks and the nature of reality,''
Bell indicated as one possibility for progress in foundations of quantum theory that ``it may be that Bohr's intuition was right---in that here
is no reality below some `classical' `macroscopic' level. Then fundamental physical theory would remain fundamentally vague, until concepts
like `macroscopic' could be made sharper than they are today'' (Bell, 1981a). 

In a comment on this paper made directly after its presentation by Bell contemporary, his friend and contemporary Abner Shimony---who, along with John Clauser, Michael Horne, and Richard Holt provided a directly experimentally testable form of Bell's inequality (Clauser, Horne, Shimony and Holt, 1969)---may have influenced Bell's thinking 
regarding measurement when he remarked that 
\begin{quote}
``perhaps I can help to focus on the source of the difficulty [in understanding Bohr's answer to E.P.R.]. In any measuring process, Bohr insists
upon a sharp distinction between object and subject. The apparatus is considered to be situated on the subject's side of this division. Hence it is 
characterized in terms of the concepts of everyday life (of which the concepts of classical physics are refinements). One may ask, however, 
whether it is possible to investigate the physical behavior of the apparatus\ldots Bohr's answer is that\ldots [it] is possible but then other apparatus
will be employed in the investigation. The boundary between the object and the subject has shifted.'' (Bell, 1981a)
\end{quote} By the time of  AM Bell was referring to this division exactly as ``the shifty split.''

The focus of the critique in AM is in fact most specifically on the conceptual imprecision involved in the treatment of {\em state evolution} during measurement that depends on the above (as Bell sees it) problematic system--apparatus division, which is typically made by having the apparatus, in one way or the other, qualify as `macroscopic.'  Bell notes that a range of different, often incompatible assumptions as to how a system can be considered to be macroscopic have been used in the standard approaches to quantum measurement (cf. Jaeger, 2014a). 
In AM, Bell analyzes the traditional treatments, which invoke sudden changes of quantum state during experiments, 
making use of the following distinction ``It will be convenient later to refer to\ldots the {\em spontaneous} jump of a macroscopic system [$S$] into a 
definite configuration, as the [Landau--Lifschitz] LL jump. And the {\em forced} jump of a quantum system as a result of `measurement' --- 
{\em external intervention} --- as the Dirac jump.'' A ``jump'' placed in the same location as the latter appears in the formulation of von Neumann; 
the postulate of state collapse according to von Neumann is also noted in the text: 
``what vN actually {\em postulates} is that `measurement' --- an
external intervention by [the rest of the world] $R$ on $S$ --- causes the state $\sum_nc_n\phi_n$ to jump, with various probabilities into
$\phi_1$ {\em or} $\phi_2$ {\em or}\ldots From the `or' here, replacing the `and', as a result of external intervention, vN infers that the 
density matrix, averaged over the several possibilities, has no interference terms between states of the system which correspond to 
different measurement results." It is this invocation of measurement at the level of postulates that Bell found extremely objectionable.

Von Neumann was clearly forced to postulate such a process, which he called ``Process 1,'' pointing out its exceptional
nature by noting that, on physical grounds, one would rather expect the more usual Process 2 to be the only one needed.
\begin{quote}
``[O]ne should expect that [Process 2] would suffice to describe the intervention caused by a measurement: Indeed, a physical intervention
can be nothing else than the temporary insertion of a certain energy coupling into the observed system, i.e., the introduction of an appropriate
time dependency of $\hat{H}$.'' (Von Neumann, 1932)
\end{quote}
Von Neumann had argued that the boundary between the measuring and measured systems should also be ``arbitrary to a very large extent'' 
(Von Neumann, 1955, p. 420) because whether the collapse happens to the measured system alone or to the joint system of measuring apparatus together
with the measured system, the statistics of outcomes will be the same from the point of view of any physical system, such as a that of a
human being, separate from them that becomes correlated with them in the same way a measurement is assumed to become during
an experiment. Although this is true, it does not aid our understanding of what takes place during measurement, but instead leaves its
details obscure. This is very clearly a case of a theory working, but working only ``FAPP.''

Bell then surveys other traditional treatments in other ``good books,'' indicating the various instances of physical imprecision within them. In the case of the 
treatment of Landau and Lifschitz (LL), which ``derive[s] the Dirac jump from the LL jump,'' he says ``In the LL formulation\ldots the theory is 
ambiguous in principle, about exactly when and exactly how the collapse occurs, about what is microscopic and what is macroscopic, what 
quantum and what classical.'' These, of course, are the most critical questions involved in a fuller understanding of the quantum mechanics
of experimentation. For this reason, imprecision in relation to these obscures the problem itself, making it all the more difficult to solve.

Bell also considers the treatment of Kurt Gottfried, and offers natural suggestions for its missing details as part of an exploration of its more
realistic character. Formally, he takes this as a treatment 
in which the density matrix $\rho$ for the joint system of system $S'=S+A$, where $A$ is the measurement apparatus system, is replaced by 
another density matrix $\hat\rho$, in which all non-diagonal elements are zero in the Hilbert space basis in which measured values and the 
apparatus `pointer' variable values are to be perfectly correlated, something which is a prerequisite of an accurate measurement often postulated by realist (as well as operationalist)
interpretations of quantum theory. To Bell, this appears to be  because there is conceptual drift ``away from the `measurement' 
(\ldots external intervention) orientation of orthodox quantum mechanics towards the idea that systems, such as $S'$ above, have intrinsic 
properties --- independently of and before observation. In particular, the readings of external apparatus are supposed to be really there before 
they are read,'' in that explication of measurement in which ``KG derives, FAPP, the LL jump from assumptions at the shifted split
$R'/S'$, which include a Dirac jump there,'' where $R'=R-A$. This is seen as having the advantage that some \begin{quote}
```macroscopic' `physical attributes'
{\em have} values at all times, with a dynamics that is related somehow to the butchering of $\rho$ into $\hat\rho$ --- which is seen as somehow
not incompatible with the internal Schr\"odinger equation of the system. Such a theory, assuming intrinsic properties, would not need
external intervention, would not need the shifty split, but the retention of the vague word `macroscopic' would reveal limited ambitions as 
regards precision.'' (Bell, 1991) \end{quote}
One might, he finally notes, avoid this term by introducing variables which have values even at small scales, as in the
deBroglie--Bohm approach. 

Bell had previously viewed the deBroglie--Bohm approach as important, in that it showed that ``the
subjectivity of the orthodox version, the necessary reference to the `observer,' could be eliminated'' (Bell, 1982). This indicates 
a potentially promising direction for increased conceptual precision, one which he sought to ``publicize.'' In this earlier discussion, he drew three 
morals from the existence of the deBroglie--Bohm model: (1) ``Always test your general reasoning against simple models,'' (2) ``in physics the only observations we must consider
are position observations, if only the positions of instrument pointers,'' and (3) one concerning terminology that was to be the main theme of ``Against measurement.'' In the paper where these morals were drawn, ``On the impossible pilot wave,'' Bell notes regarding (3) that ``serious people'' were likely ``misled by the pernicious misuse of the word `measurement''' which ``strongly suggests the ascertaining of some preexisting property of some thing, any instrument
involved playing a purely passive role. Quantum experiments are just not like that, as we learned especially from Bohr'' (Bell, 1982).

However, as seen below, by the end of the 1980s, Bell found a different approach more promising, one that deviates from standard
quantum mechanics at the level of law: Indeed, he had concluded already concluded by the time of his article ``Are there quantum jumps?'' that ``If, with 
Schr\"odinger, we reject extra variables, then we must allow that his equation is not always right\ldots it seems to me inescapable\ldots a recent idea [of Ghirardi, Rimini, and Weber (Ghirardi, Rimini and Weber, 1985)], a specific form of spontaneous wavefunction collapse, is particularly simple and effective'' (Bell, 1987).

\section{Modified Quantum Dynamics}

\subsection{The Desiderata and Superpositions at Large Scales}

In addition to their common use of notions of measurement and macroscopicity, standard analyses of  quantum mechanical
situations, as viewed from the perspective of data production, suffer from what has been called the ``opportunistic employment of the superposition principle.'' That is, one is tempted to allow the superposition principle to operate whenever convenient and not operate whenever inconvenient,
as opposed to understanding via specific basic quantities precisely when it may or may not be in force. This issue was taken up by Shimony in 
his article ``Desiderata for a modified quantum dynamics,'' presented in a memorial session for Bell, wherein he also noted that 
``At a workshop at Amherst College in June {\em Bell} remarked that the stochastic modification of quantum dynamics is the most important 
new idea in the field of foundations of quantum mechanics during his professional lifetime'' (Shimony, 1991).

It is clear in AM that Bell viewed the modification of the standard quantum dynamics and presence or absence of state
superposition as important, in that it is a move that
provides an opportunity to correct at least some of the forms of imprecision noted above by providing objectivity to the circumstances 
under which measurement-like events would or would not take place. It is, therefore, worth looking more closely at the context in which 
such theories can be developed. This is just want  Shimony does in ``Desiderata...'', by spelling out four assumptions 
``concerning the interpretation of the quantum mechanical formalism have the consequence of making the [measurement problem and the problem 
of Schr\"odinger's cat] so serious that it is difficult to envisage their solution without some modification of the formalism itself.'' 
These assumptions, variously sanctioned in AM,  are generally ``strongly supported by physical and philosophical considerations, and therefore a high price would be paid by sacrificing one of them in order to hedge standard quantum mechanics against modifications.'' They are the following (Shimony, 1991). 

\begin{quote} 
(i)  ``The quantum state of a physical system is an objective characterization of it.'' (As Bell puts it in AM, a ``serious
formulation will not exclude the big world outside the laboratory'' and will not be concerned exclusively with ``piddling laboratory operations.'')
\end{quote}

\begin{quote} 
(ii) Connected with Bell's theorem: ``The objective characterization of a physical system by its quantum state is complete, 
so that an ensemble of systems described by the same quantum state is homogeneous, without any differentiations stemming from differences 
in `hidden variables.'\ \hskip -2pt'' 
\end{quote}

\begin{quote} 
(iii) ``Quantum mechanics is the correct framework theory for all physical systems, macroscopic as well as microscopic, and hence it specifically applies to measuring apparatuses.'' (About which, however, it should be noted that ``The main consideration 
in favor of [it being] the incompatibility proved by Bell (1987, pp.14-21 and 29-39) between quantum mechanics and local hidden variables 
theories, but Bell himself emphasizes that there is still an option of non-local hidden variables theories, which he does not regard as 
completely repugnant.'' Furthermore, this assumption has in Bell's eyes the implication that all variants of the Copenhagen interpretation are ``ruled out.'')
\end{quote}

\begin{quote} 
(iv) ``At the conclusion of the physical stages of a measurement (and hence, specifically, before the mind of an 
observer is affected), a definite result occurs from among all those possible outcomes (potentialities) compatible with the initial state of the 
object.'' (Bell is skeptical even of having {\em biology} pertinent to measurement induction---hence his comment regarding collapse 
``Was the wavefunction of the world waiting to jump for thousands of years until a single-celled living creature appeared\ldots or some better qualified system \ldots with a PhD?'' (Bell, 1981b)
\end{quote}

\noindent Shimony then sets out a list of eight well supported desiderata for such a dynamics, the last pertaining critically
to the proposal Bell looked to, namely, that of GRW. 
\begin{quote}
``The modified dynamics should be capable of accounting for the occurrence 
of definite outcomes of measurements performed with actual apparatus, not just with idealized models of apparatus. The Spontaneous 
Localization theory of [GRW '86] has been criticized for not satisfying this desideratum\ldots Albert and Vaidman
(Albert 1990, 156-8) note that the typical reaction of a measuring apparatus in practice is a burst of fluorescent radiation, or a pulse of 
voltage or current, and these are hard to subsume under the scheme of measurement of the Spontaneous Localization theory.'' (Shimony, 1991)\end{quote}
Shimony also notes difficulties in this approach and others pertaining to Bell's concern about {\em irreversibility} (which Bell says in AM also defies a precise conceptual basis): ``a stochastic modification of quantum dynamics can hardly avoid introducing time-asymmetry. Consequently, 
it offers an explanation at the level of fundamental processes for the general phenomenon of irreversibility, instead of attempting to 
derive irreversibility from some aspect of complexity (which has the danger of confusing epistemological and ontological issues).''

Now, Bell was not the first to notice the imprecision of the traditional approach to quantum measurement. Notably, Wigner---who
had shown the limitations of von Neumann's arbitrariness of the location of the division involved in his own measurement schema by
showing that if a cognitive systems is used as a measurement apparatus contradictions can appear (Wigner, 1963)---pointed
this out very clearly in his critique of one highly developed standard treatment: In what was perhaps the most sophisticated treatment 
within that approach, that of Danieri, Loingier and Prosperi, the three authors were said by him to be ``using phrases such as `macroscopic 
variables' and `macroscopic objects' without giving a precise definition of these terms'' (Freire, 2005) so that, for example, their premisses could 
not be rigorously formulated. 
In remarks on Prosperi's paper in a key meeting of measurement theorists at the outset of the 1970s, which continued along the lines of 
the DLP approach, Wigner noted specifically the inappropriateness of making use of ``something as inadequately defined as is the macroscopic 
nature of something'' in serious physical discussions (his remarks immediately follow the article of Prosperi (1971)). He noted that ``the theory of the interaction of a quantum system with a classical (macroscopic) system has not been formulated so that 
the mathematical meaning of the arrows [indicating the change of joint-system state-vector upon measurement] is not clear'' (Wigner, 1971, p. 7). \begin{quote}
``[M]ost quantities which we believe to be able to measure, and surely all the very important quantities such as position, momentum, fail to commute with all the conserved quantities so that their measurement cannot be possible with a microscopic apparatus. This raises the suspicion 
that the macroscopic nature of the apparatus is necessary in principle and reminds us that our doubts concerning the validity of the superposition 
principle for the measurement process were connected with the macroscopic nature of the apparatus.'' (Wigner, 1971)
\end{quote}
And this nature, now sometimes referred to by the term `macroscopicity', is not rigorously characterized within that approach.

In the years between Wigner's critique and Bell's later criticisms of traditional quantum measurement theory, Anthony Leggett had considered 
performing tests for the quantum effects in ``macroscopic systems,'' preferrably large material systems, to better illuminate the question of whether there is a clear role for macroscopicity in measurement. Leggett still wishes to find ``evidence of a breakdown of the quantum mechanical scheme of the physical world [in] that 
which connects the world of atoms and electrons, for which it was originally developed, with the `everyday' world of our immediate experience,'' 
where quantum mechanically complementary properties appear compatible (Leggett, 2002). In particular, he wishes to find superposition and, so, interference effects DLP had argued should not occur during measurements (cf. Home and Whittaker, 2002). 
For this purpose, Leggett has suggested studying superconducting devices (SQUIDs) and the Josephson effect, in which states of a current of electrons could, in principle, enter a superposition of states of clockwise and/or anti-clockwise circulation (Leggett, 2000).  
Since the early work of Leggett,  `macroscopic' has been increasingly defined in terms 
large values of specific observable quantities, generalizing Bohr's original belief that heft and rigidity andx that of others simply that a sufficiently large number of degrees of freedom are essential,  rather than being identified via criteria related to the resolution of naked eye, as the most direct understanding of the meaning of the term would suggest (Jaeger, 2015). 

Leggett has proposed a measure he calls disconnectivity 
$D$, a ``semi-quantitative'' and ``qualitatively defined''  notion, claiming that ``the quantum states important in the discussion of the [cat] paradox 
are characterized by a very high value of \ldots `disconnectivity'; by contrast, the states necessary to explain so-called `macroscopic quantum phenomena' in superfluids and superconductivity have only low disconnectivity, so that they are irrelevant to our question\ldots'' (Leggett, 1980).
Rather, these center on the ``most promising area to look [for high disconnectivity states is that of] phenomena where quantum tunneling plays an essential role'' (Leggett, 1980). The GRW approach can be viewed as having participated in this trend as well.

\subsection{Continuous Spontaneous Localization and Beables}

It was the direction of GRW, shared by other workers such as Philip Pearle and conveniently called
continuous spontaneous localization (CSL) that Bell saw, in the period in which AM appeared, as most clearly offering an alternative and 
``explicit model allowing a unified description of microscopic and macroscopic systems.'' The starting point is one of ``a modified quantum dynamics 
for the description of macroscopic objects'' in which systems of many components have wave-functions that frequently spontaneously localize to small regions, claiming that with it ``most features of the behavior of macroscopic objects are accounted for by quantum mechanics in a natural way, due to the irrelevant spreads of wave packets for macroscopic masses" (Ghirardi, Rimini and Weber, 1985). 

The central parameters of interest in this model were laid out by GRW as follows.
``If one assumes for simplicity that the localization frequencies $\lambda_i$ of all microscopic (e.g., atomic) 
constituents of a macroscopic body are of the same magnitude\ldots, the center of  mass is affected by the 
same process with a  frequency $\lambda_{\rm macro}=N\lambda_{\rm micro}$ \ldots where [the ``macroscopic number''] $N$ is of the order of Avogadro's number" (Ghirardi, Rimini and Weber, 1985). 
In response to this work, Bell noted that
\begin{quote}
``In the GRW scheme this vagueness [regarding wavefunction collapse]  is replaced by mathematical precision. \ldots 
departures of the Schr\"odinger equation show up very rarely and very weakly in few-particle systems. But in macroscopic systems, 
{\em as a consequence of the prescribed equations,} pointers very rapidly point, and cats are very quickly killed {\em or} spared.''
(Bell, 1991). 
\end{quote}
Bell saw this aspect of the GRW approach in marked, positive contrast to examples of `solutions' of the measurement
problem involving infinite limits that had appeared, in particular, the 1972 model of Coleman and Hepp, where a solution for the dynamics 
of a model apparatus consisting of a semi-infinite array of spin-1/2 particles was given, which was viewed by some as a sort of solution 
to the measurement problem. Bell was critical of the Coleman--Hepp model, noting that, for it, ``the rigorous reduction does not occur in physical time but only 
in an unattainable mathematical limit\ldots the distinction is an important one'' (Bell, 1975). 

One difficulty subsequently encountered by the CSL approach is finding a set of parameters that allow 
it to describe what is observed. This difficulty is connected with what Bell called ``beables,'' those quantities which could be understood
realistically and which could correspond with what is actually observed; those beables associated with local space-time regions are ``local beables.''
Bell viewed contemporary quantum mechanics textbooks as failing to focus on these quantities.
\begin{quote}
``What you may find there are the so-called `local observables'. It is then implicit that the apparatus of the `observation', or, better, of experimentation,
and the experimental results are real and localized. We will have to do the best we can with these rather ill-defined local beables, while hoping
always for a more serious reformulation of quantum mechanics where the local beables are explicit and mathematical rather than
implicit and vague'' (Bell, 1990)  
\end{quote}
For their part, CSL theories have mainly followed what has been called the mass density ontology (Allori, Goldstein, Tumulka and Zhang\'i, 2008), as evidenced, for example, by the use of the parameter $N$ above, to which mass density would be proportional for systems built from a given sort of fundamental subsystem. For a system in a superposition of states with differing mass densities for which there is an operator, the larger the difference 
of the mass density distribution of the states is, the more quickly a collapse will take place. Thus, the collapse rate for superpositions states of microscopic systems is low because the mass density differences are likewise low, and for superpositions of macroscopic states it is large because the mass density differences are likewise large.

However, there is a problem of persistent ``tails'' for any collapse process that completes in finite time: State
functions correspondence to perfectly sharp values in position  not obtained in finite time in cases where sharp values obtained and it
is admitted that ``[f]or a macrosystem, the precisely applied eigenstate--eigenvalue link does not work'' (Pearle, 2009). At least one of
the longest and most active advocates of CSL, Pearle does not see this as precluding the success of the theory, despite 
Shimony's desideratum which regards tails specifically, namely,
\begin{quote}
``d. If a stochastic dynamical theory is used to account for the outcome of a measurement, it should not permit excessive indefiniteness of the outcome, where ``excessive" is defined by considerations of sensory discrimination. This desideratum tolerates outcomes in which the apparatus variable does not have a sharp point value, but it does not tolerate `tails' which are so broad that different parts of the range of the variable can be discriminated by the senses, even if very low probability amplitude is assigned to the tail. The reason for this intolerance is implicit in Assumption (iv)\ldots If registration on the consciousness of the observer of the measurement outcome is more precise than the `tail' indicates, then the physical part of the measurement process would not yield a satisfactory reduction of the initial superposition, and a part of the task of reducing the superposition would thereby be assigned to the mind. For this reason, I do not share the acquiescence to broad `tails' that Pearle advocates (1990, pp. 203-4)\dots'' (Shimony, 1991)
\end{quote}

Pearle has argued more recently, as he had once in Bell's presence---at the same Amherst conference mentioned above---that ``one should not express a new theory in an old theory's language,'' a comment ``at which he beamed'' (Pearle, 2009). In
particular, Pearle argues that ``a collapse theory is different from standard quantum theory and\ldots therefore requires a new language, conceptual
as well as terminological'' (Pearle). Emphasis is put, for example, on ``near possessed'' rather than ``possessed'' values of physical quantities.
In his explication of CSL, Pearle argues that  
\begin{quote}
``CSL retains the classical notion that the physical state of a system corresponds to the state vector. Corresponding to a random field $w({\bf x},t)$ whose probability of occurrence is non-negligible, the dynamics always evolves a realizable state. Therefore, one is freed from requiring the (near) eigenstate-eigenvalue link criterion for the purpose of selecting the realizable states. I suggest that the eigenstate-eigenvalue link criterion be subsumed by a broader concept. It must be emphasized that this new conceptual structure is only applicable for a theory which hands you macroscopically sensible realizable states, not superpositions of such states. In the new language, corresponding to a quantum state, every variable possesses a distribution of values\ldots'' (Pearle, 2009)
\end{quote}
The notions of this new `language' are to be given meaning by considering ways in which it is to be used in context.
The distribution here is not to be understood as a probability distribution, despite possessing all the defining properties of one, because, argues Pearle, it is unlike classical physics where probabilities are understood as due to {\em ignorance}. Here,
 one
\begin{quote}
``may give the name `stuff' to a distribution's numerical magnitude at each value of the variable, as a generalization of Bell's quasi-biblical characterization, `In the beginning, Schr\"odinger tried to interpret his wavefunction as giving somehow the density of the stuff of which the world was made.' One is encouraged to think of each variable's stuff distribution as something that is physically real. The notion allows retention of the classical idea that, for a physical state, every variable possesses an entity. What is different from classical ideas is that the entity is not a number.'' (Pearle, 2009)
\end{quote}
(Note that Bell had used the term `stuff' in the context of stochastic localization theory in AM, as follows.
\begin{quote}
``The GRW-type theories have nothing in their 
kinematics but wavefunctions. It gives the density (in a multi-dimensional configuration space!) to {\em stuff}. To account for the narrowness
of that stuff in macroscopic dimensions, the linear Schr\"odinger equation has to be modified, in the GRW picture by a mathematically
prescribed spontaneous collapse mechanism.'' (Bell, 1991) )
\end{quote}
On this view, {\em every variable} possesses such a distribution, so that ``complementarity here means that variables whose operators do not commute 
do not possess joint distributions, but they do jointly possess distributions'' (Pearle, 2009).  As an example, one can consider a state that is the quantum superposition of two, one with state amplitude $\sqrt{1-\epsilon}$ and the other with amplitude $\sqrt{\epsilon}$. Under the above interpretation,
the smaller ``tail'' state is considered to represent ``an unobservably small amount of stuff which allows describing the state vector by (qualified)
possessed values assigned to macroscopic variables, consistent with the dominant state'' (Pearle, 2009). 

As noted above, a central problem for CSL is finding parameter ranges for which it would have experimental predictions deviating from those
of standard quantum mechanics. Interference experiments are archetypical and could serve to differentiate the two, because CSL tends to
destroy interference in that it naturally destroys one of the necessary pair distinct states in the case of massive systems, with an interference
visibility that decreases with the increase in system mass. One might test the theory by considering, for example, two-slit experiments on 
each of a range of sorts of systems differing in their masses: photons, electrons, neutrons, atoms, and molecules. At one limit of this range,
one has the photons, of course, are massless, and CSL would have no additional effect. In the upper range of the experiments that have
been performed, one finds the C-60 molecule, which has $N=720$ nucleons. 
The value 720 is too small for a great impact, and is very much smaller than the Avagadro number often taken as a value one can say is clearly ``macroscopic.'' Because collapse narrows wavepackets, it also leads to a momentum increase and hence to an energy increase, requiring collapse rates that differ not only with particle number, but also particle mass (Pearle, 2009). 
Unfortunately, experiments capable of testing this hypothesis are not of the sort commonly performed and currently await testing. 

\section{Toward the Removal of Conceptual Imprecision}

In the quantum theory of measurement, experiments are typically understood schematically as follows. A system S is initially prepared in a quantum 
state $|\Psi\rangle$ through a series of physical interactions, after which it is measured through interaction with an apparatus A which is required, in the process,
to enter a state the value of the ``pointer'' property which itself becomes perfectly correlated with the value of the measured property $E$ of S. 
A minimal requirement placed on a measurement is that a certain ``calibration condition'' be satisfied, namely, 
that if a property to be measured is a real one, then it should exhibit its value unambiguously and with certainty, cf. (Busch, Lahti and Mittelstaedt, p. 28). 
For so-called ``sharp observables,'' that is, properties represented by Hermitian operators, this calibration condition is equivalent to a probability reproducibility condition, namely, that a probability measure $E_{\Psi}$ for a property be ``transcribed'' onto that of the corresponding apparatus pointer property. Measurement is also taken to include the reading out of the registered value in addition to the above process of registration of the measured property by apparatus A. 
The question of how this pointer ``objectification'' is achieved in view of the nonobjectivity of the measured operator, is the first part of the so-called ``objectification problem.''  

The second part of the objectivication problem is that of ``value objectification.'' A pointer reading refers to the property 
value of the object system prior to measurement only if the measured observable was objective {\em before} the measurement. When the observable
is non-objective, the question arises what happens to the system in the course of the measurement. In general, some state change is unavoidable. The attempts to minimize this irreducible `disturbance' then naturally lead to the  concept of ideality of a measurement. Ideality requires another characteristic, namely, repeatability:
A {\em repeatable} measurement will put the system in a state in which the pointer reading X refers to an objective value of the measured  observable. 
This is taken to show that the existence of repeatable measurements is necessary for realistic interpretations of quantum mechanics (Busch and Jaeger, 2010). 
For such measurements, pointer objectification entails value objectification via a strong value correlation. Such an operational approach to measurement, however, threatens to mask the objective, physical nature of measurements in themselves with which Bell was so concerned. 

Already in his 1981 article ``Quantum theory for cosmologists,'' Bell asked the following rhetorical questions about quantum measurements as understood within such a scheme.
\begin{quote}
``If [quantum] theory is to apply to anything but idealized laboratory operations, are we not obliged to admit that more or less `measurement-like' 
processes are going on more or less all the time more or less everywhere? \ldots 
The concept of `measurement' becomes so fuzzy that it is quite surprising to have it appearing in physical theory at the most fundamental level\ldots 
[D]oes not any analysis of measurement require concepts more fundamental than measurement? And should not the fundamental theory be about these...?"
(Bell, 1981b)
\end{quote}
There is a stark contrast between the everyday ``classical'' measurements and quantum measurements. 
In classical physics, it is certainly the case that situations that are more or less measurement 
like are going on all the time everywhere. The difficulty in the quantum case is that, although measurement processes should be similarly happening, the outcomes of careful experiments found by us are consistent with the predictions made using the Schr\"odinger state evolution, even though the superposition principle should not apply when such measurement-like processes are taking place, if human beings are to be treated just like other physical entities.

If it is indeed the case that measurements, as distinct from subjective acts of observation, are nonetheless an integral part of physics and 
{\em not} artificially introduced, the special physical circumstances appearing in measurements must be circumscribed. In the search for physical clarity, 
we can remove anthropocentric elements from our conception of quantum measurement by finding the set of radically influential objects corresponding to this kind rather than a generic {\em apparatus} for measurements, and thus remove impediments to progress in isolating the physical conditions underlying measurement as an objective process. If human beings or other, larger sets of, for example, biological entities 
precipitate such physical conditions then the special role of these entities will have been objectively grounded and natural philosophy will have been advanced.

It may be helpful to consider the possibility that the set of related entities as being of a {\em natural kind}, because when successful, classes employed by science do correspond to 
natural kinds, as in the cases of the sorts of chemical element, subatomic particle, star, and galaxy (Jaeger, 2014b). This does not require that these objects be treated as entirely different from other physical objects, but only assists in our comprehending the implications of their set of common characteristics.
In addition to having particular sets of natural properties in common with one another, these tokens should be subject to laws of nature relevant to these properties. 
One can seek a set of conditions for being a member of the kind `radically influential object' as a way of making progress
toward an improved realist physics. From the formal point of view, such influencers are typically assumed to have these characteristics 
in common: they i) induce non-unitary state-change and 
ii) satisfy the conditions on the systems for providing a robust record of measurement outcomes. 
The former relates to the ability to disentangle the joint state of the influencer--target system and the latter corresponds to the 
production of Einsteinian elements of reality. However, as Bell rightly pointed out, more fundamental properties
should be present, of which these properties are {\em consequences}. It is these configurations of fundamental properties that should play a key role in describing measurement via truly fundamental physical principles.

Some models of measurement, discussed further below, assume that measurements involve complex measuring systems with a large number of degrees 
of freedom prepared in metastable states. Some natural systems are known to us to measure---for example, our eyes when connected with our nervous system are such systems.   
Consider the human optical system taken to comprise all the material from the eyes to brain, asking whether previously assumed characteristics are essential.  We must look to our understanding of the behavior of macromolecules of the optical nervous system when light is incident upon it. Shimony pointed out that the photoreceptor protein of the rod cells, known as rhodopsin, absorbs photon followed by a biochemical cascade that is then followed by an electrical pulse in the optic nerve. Shimony notes that rhodopsin has two components, 
``retinal, which can absorb a photon, and opsin, which acts as an enzyme that effects the binding of about five hundred mediating molecules when it is triggered by the excited retinal,'' and asks
\begin{quote}
[W]hat if the unitary dynamics of evolution of the photon and the retinal produces a superposition of the cis and the trans conformations? ...Would not such a superposition produce an indefiniteness of seeing or not seeing a visual flash, unless, of course, a reduction occurred further along the pathway from the optic nerve to the brain to the psyche?'' (Shimony, 1991) 
\end{quote}
The distinct, alternative physical states corresponding to different conformations of a molecule which can superpose and then enter a specific state when in contact with the remainder of the nervous system are  central to the functioning of this light detection process in a biochemical and electric realm. The presence of this larger subsystem beyond rhodopsin has an effect of ``amplification.'' Some artificial systems can also mimic the behavior of such natural radical influencers, and so are exploitable by designers of 
experiments, e.g. avalanche photodiodes (APDs) plus electronics. These systems, for example, are rather complex and involve many 
degrees of freedom and metastable initial states. Their effects appears to us to be completed in a way that, for example,
a  Stern--Gerlach magnet alone without a downstream beam-occupation detector are not. 
Nonetheless, it has been argued by some, like Asher Peres (Peres, 1980), that amplication is not required for measurement to take place, an argument considered below in greater detail. 

It is the resolution of questions  such as these regarding requirements that could provide the characteristics of the measurers as a natural kind and would be a helpful element of any realist treatment of quantum state change, such as that called for by Bell. Although, like Bell, I am concerned about the notion of measurement being given an unusually prominent place in physics, I am much more concerned about notions associated with other terms that he criticizes in less detail in ``Against measurement,'' such as `observable' and `observation,' which are clearly laden with the influence of theory and importantly with a directly {\em subjective}, that is, non-physical aspect. This should be kept in mind when asking the question of which characteristics of the above example are {\em necessary} to an objective understanding of measurement in contrast to those which may only be systematically present in the considerations of physicists only because they themselves are subjects. 
Although at least one human knower is always {\em eventually} present who is witness in any successful experiment---or, in less contrived cases, simply observes ongoing natural events---who happens also to be large, this should not be allowed to beg the question of whether largeness is a necessary characteristic of a radical influencer in all data-yielding situations.
 
The matter of amplification, which aids in the delivery of signals perceptible to the human senses, is similarly subtle.
Amplification has often been considered to lead to the objectivity of data produced because, from the statistical point of view, it makes it is very difficult to undo and allows many systems and observers to be affected by the resulting signal. 
Despite the presence of this and other characteristics in many or most familiar situations wherein one learns about the state of a part of the universe, one must ask whether there is an objective reason for requiring them in data-yielding situations {\em in general}. 
Another characteristic mentioned above that is often selected out for special status is {\em complexity}. Notably, it was indicated by DLP, who required in their treatment that interference ``be absent by virtue of the complexity of the considered system,'' with complexity taken to refer to the number of degrees of freedom of the joint system being large (Danieri, Loinger and Prosperi, 1962). It is also strongly indicated as a requirement in the above mentioned work of Peres that ``benefited from comments by J.S. Bell'' entitled ``Can we undo quantum measurements?'' (Peres, 1980). Peres introduces a simple measurement model and with it demonstrates that systems of many degrees of freedom may effectively obey superselection rules because it is impossible to measure any phase relationship between two quantum states in the limit of an {\em infinite number} of degrees of freedom. He does does so without requiring another common candidate requirement, namely, a measurer prepared in metastable state. The demonstration is a sensible one FAPP, but only FAPP, like many others. Its value is that it shows that amplification and the presence of metastable states of the measuring system are {\em not} necessary for measurement.

All the commonly assumed characteristics of measurers might be thought by us to be natural simply because they reflect a subtle and unnoticed anthropocentrism, seeming natural to us only because we scientists are human. Just because humans are comparative large physical systems and perform measurements in the experimental context does not mean that we must have access to all measurement results:  measurement and the experience of a measurement are distinct, despite their going together in our human experience. As Bell suggests, must ask ourselves what the purely objective properties of measurement processes might be that could be solely responsible for, or most significant to the occurrence of every measurement. 
We see above, for example, that having entered the field of a Stern--Gerlach magnet alone is insufficient for the measurement of an appropriate particle's spin and that, following its action, there is a sufficient set of elements present for a successful measurement to take place when detector plus electronics capture the output beams which can later be viewed by an experimenter. What is happening in this case? According to the Schr\"odinger evolution, the effect of the magnetic field is only to entangle the particle's spin with its direction; it is usually understood that it is the detector suite that allows the spin to be identified after the above spin--path correlation has occurred. The detector, unlike the magnetic field created by the magnet, is complex, as is the human nervous system. 

The one characteristic that appears to survive our removal of unnecessary conditions is what appears to be something similar to physical  complexity. It is helpful now to recall, as Shimony noted, that a benefit of stochastic modification of quantum dynamics was to offer an explanation (``for the general phenomenon of irreversibility'') at the level of fundamental processes, something clearly in harmony with Bell's call for a  notion more fundamental than measurement to account for the emergence of experimental data. We should, of course, also note the warning that looking to ``some aspect of complexity'' has ``the danger of confusing epistemological and ontological issues.'' 
Proceeding cautiously, then, we can ask in precisely what sense the ontology of above situations might be complex.

The commonly recognized measurement-like situations not only involve a significant number of degrees of freedom but may involve distinct parts. (Incidentally, the notion of natural kinds can also assist in distinguishing between internal and external degrees of freedom, via the notion of fundamental entities.)
An important, well known example relevant to our considerations here is the Schr\"odinger cat experiment. In this thought experiment, 
Schr\"odinger considered an unstable atom A the decay of which would release a hammer H that, would break a vial containing poison and allowing the cat C, which is taken as otherwise isolated from the rest of the universe, to be exposed to the poison (Schr\"odinger, 1935). 
Aside from the ``absurdity'' (as he calls it) of the appearance there of two equiprobable distinct states of the cat's `health' at a particular moment in the corresponding 
overall state
\begin{equation}
|\Psi\rangle ={1\sqrt{2}}(|{\rm undecayed}\rangle_{\rm A} |{\rm unreleased}\rangle_{\rm H} |{\rm alive}\rangle_{\rm C}+
|{\rm decayed}\rangle_{\rm A} |{\rm released}\rangle_{\rm H} |{\rm dead}\rangle_{\rm C}) 
\end{equation} 
of the joint system, this situation has been viewed (in the terminology to be replaced in a more precise conception of measurement) as linking the atom to the `macroscopic' domain (cf. Jaeger, 2014a). One can equally well, as Bell's position would relatively favor, view this as a situation involving complex physical circumstances. Clearly, the joint system of A+H+C is more complex in several respects than that of C taken alone---for example, it has more subsystems, more degrees of freedom, and involves an interaction of C with H. The same can be said for A+H relative to A+H+C, which although A+H already involves an interaction, involves additional interactions. But, C itself involves internal interactions. One can ask, for example, whether these overlooked internal interactions are significant in relation to the complexity of the situation or otherwise influence measurement, because no-one has ever observed that sort of alive--dead circumstances in a cat or similar being.

Finally, although the above example continues to be of conceptual importance, one is now in need of further, and more practical examples to make progress. On the side of practical, rather than thought experiments, let us consider as an example of what is now available to be explored, some experiments performed relatively recently by Gerlich et al. (Gerlich, Eibenberger, Tormandi et al., 2011). These experiments have been taken by their creators to ``prove the quantum wave nature and delocalization of compounds composed of up to 430 atoms, with a maximal 
size of up to 60 {\AA}ngstroms, masses up to $m = 6,910$ AMU.'' To us, they can be seen to involve something somewhat similar to the suprising predictions of quantum mechanics in the Schr\"odinger cat thought experiment. These quantum systems involve thousands of internal degrees of freedom. 

An important point here is the factor in these experiments that differentiates the detector used from conventional detectors operating in the visible and near infrared, such as avalanche photodiodes and photomultiplier tubes of past quantum optics experiments, namely, that although those may be single-photon sensitive, they could not reliably determine the {\em number} of photons in a pulse of light like the photon-number-resolving detectors used in this experiment. Such deterimination is made possible as follows. The experiments used
a calorimetry-based photon detector in which energy is deposited in an absorber whose thermometer was determined
via an observed change in temperature. Tungsten transition-edge sensors were used and understood to operate in such a way that tungsten
electrons act as both energy absorber and thermometer, and were prepared as to keep the tungsten electrons on the edge of a superconducting-to-normal-conduction transition; a dependence of resistance on temperature was set up so as to allow precise thermometry. 
The change of current in the voltage-biased detector was measured with a superconducting quantum-interference device (SQUID) array and analyzed.
The results demonstrated quantum state superpositions of states of certain properties of large entities, suggesting that the  number of degrees of freedom involved does not correspond to the sort of physical complexity required for induced behavior which would differ from that given by the standard quantum mechanical desciption. 

The detailed study of such experiments, in particular, the array of instrumentation and its necessity for successfully providing data describing quantum phenomena, alongside the study of human sensory systems themselves, can be expected to yield a more refined understanding of 
measurement-like processes within quantum theory and should provide novel insights allowing more precise vocabulary and concepts to be introduced to improve upon its basic principles, as John Bell recommended.
Novel approaches to measurement in quantum optics, much as in the testing of the Bell--CSHS inequality itself in the past, can also aid us in transcendending the current limitations illustrated by the weakness which he identified in quantum physical terminology deployed in measurement situations. After such work, one will be in a better position to consider specific modifications of the fundamental laws of quantum physics or the quantum state description itself, both of which should help us progress in the direction that Bell suggested we go in order to advance natural philosophy.

\vfill\eject

\noindent {\bf References}
\\

\noindent Albert, D. (1990). On the collapse of the wavefunction. In A. I. Miller (ed.) {\em Sixty-Two Years of Uncertainty: Historical, philosophical, and physical inquiries into the foundations of quantum mechanics}. New York: Plenum Press, 
p. 153-165.\\

\noindent Allori, V., Goldstein, S., Tumulka, R. and Zanghi, N. (2008). On the common structure of Bohmian mechanics and the Ghirardi-Rimini-Weber theory. British Journal for the Philosophy of Science {\bf 59}(3), 353-389.\\

\noindent Bell, J.S. (1990). Against measurement. In A. I. Miller (ed.) {\em Sixty-Two Years of Uncertainty: Historical, philosophical, and physical inquiries into the foundations of quantum mechanics}. New York: Plenum Press, 
pp. 17-31.\\

\noindent Bell, J.S. (1981a). Bertlmann's socks and the nature of 
reality. Journal of Physics {\bf 42}, Colloque C2, suppl\'ement au No. 3, p. 41.\\

\noindent Bell, J.S. (1981b). Quantum mechanics for cosmologists. In C. Isham, R. Penrose and D. Sciama (eds.) {\em Quantum Gravity 2}. Oxford: Clarendon Press, pp. 611-637.\\

\noindent  Bell, J.S. (1975). On wave packet reduction in the Coleman--Hepp model. Helvetica Physica Acta  {\bf 48}, 93-98.\\

\noindent Bell, J. S. (1990) La nouvelle cuisine. In  {\em Between Science and Technology} A. Sarlemijn and P. Kroes (eds.).
Dordrecht: Elsevier Science Publishers.\\

\noindent Busch, P., Lahti, P. and Mittelstaedt, P. (1991). {\em The Quantum Theory of Measurement}. Heidelberg: Springer.
\\

\noindent Busch, P. and Jaeger, G. (2010). Unsharp Quantum Reality. Foundations of Physics {\bf 40}, 1341. \\

\noindent Clauser, J., Horne, M., Shimony, A. and Holt, G. (1969). Proposed experiment to test local hidden-variable theories. Physical Review Letters {\bf 23}, 880-884.\\

\noindent Danieri, A., Loinger, A. and Prosperi, G.M. (1962). Quantum theory of measurement and ergodicity conditions. Nuclear Physics {\bf 33}, 297-319.\\

\noindent Freire, O. (2005). Orthodoxy and heterodoxy in the research on the foundations of quantum mechanics: E. P. Wigner's case. In B. S. Santos (ed.) {\em Cognitive Justice in a Global World}. Madison: University of Wisconsin Press.\\

\noindent Gerlich, S., Eibenberger, S., Tomandi, M. et al. (2011). Quantum interference of large organic molecules. Nature Communications {\bf 2}, 263.\\

\noindent Ghirardi, G. C., Rimini, A. and Weber, T. (1986). Unified dynamics for microscopic and macroscopic systems. Physical Review D {\bf 34}, 470-491.\\

\noindent Home, D. and Whitaker, A. (2007). {\em Einstein's Struggles with Quantum Theory: A reappraisal}. New York: Springer, p. 310.\\

\noindent Jaeger, G. S. (2014a). What in the (quantum) world is macroscopic? American Journal of Physics {\bf 82}, 896-905.\\

\noindent Jaeger, G. S. (2014b). Macroscopic realism and quantum measurement: Measurers as a natural kind"
Physica Scripta {\bf T163}, 014017.\\

\noindent Leggett, A.  (1980). Macroscopic quantum systems and the quantum theory of measurement. Progress in Theoretical Physics Supplement {\bf 69}, pp. 80-100.\\

\noindent Leggett, A. (2000). The order parameter as a macroscopic wavefunction. In J. Berger and J. Rubinstein (eds.) Lecture Notes in Physics m62. Heidelberg: Springer, pp. 230-238.\\

\noindent Leggett, A. (2002). Testing the limits of quantum mechanics: Motivation, state of play, prospects. Journal of Physics: Condensed Matter {\bf 14}, R415-R451.\\

\noindent  Pearle, P. (2009). How Stands Collapse II. In  W. Myrvold and J. Christian (eds.)  {\em Quantum Reality, Relativistic Causality, and Closing the Epistemic Circle} (The Western Ontario Series in Philosophy of Science, Vol. 73). New York: Springer, pp. 257-292.\\

\noindent Peres, A. (1980). Can we undo quantum measurements? Physical Review D {\bf 69}, 879-883.\\

\noindent Prosperi, G. M. (1971). Macroscopic physics and the problem of measurement. In {\em Foundations of Quantum Mechanics. Proceedings of the International School of Physics ``Enrico Fermi''} London: Academic Press, pp. 97-126.\\

\noindent Von Neumann, J. (1932). {\em Mathematische Grundlagen der Quantenmechanik}. Berlin: Julius Springer.\\

\noindent Von Neumann, J. (1955). {\em Mathematical Foundations of Quantum Mechanics}. Princeton, NJ: Princeton University Press, Ch. V, Sect. 4.\\

\noindent Wigner, E.P. (1963). The problem of measurement. American Journal of Physics {\bf 131}, 6.\\

\noindent Wigner, E.P. (1971). The subject of our discussions. in {\em Foundations of quantum mechanics. Proceedings of the International School of Physics ``Enrico Fermi.''} London: Academic Press, p. 5.\\

\noindent Schr\"odinger, E. (1935). Die gegenw\"artige Situation in der Quantenmechanik. Die Naturwissenschaften {\bf 23}, 807, 823, 844.\\

\vskip 30pt

\noindent Acknowledgment of Support. This work was partly supported by funding from DARPA QUINESS program through U.S. ARO Award W31P4Q-12-1-0015. 

  \cleardoublepage

\end{document}